\documentclass[aps,prl,twocolumn,preprintnumbers]{revtex4-2}
\usepackage{amsmath, mathrsfs, amssymb,amsfonts,amsthm,graphicx, epsf, dcolumn, yfonts}
\usepackage[hyperfootnotes=true]{hyperref}
\usepackage{color}
\usepackage{slashed}
\usepackage{setspace}
\usepackage{cancel}
\usepackage{wasysym}
\usepackage[lofdepth,lotdepth]{subfig}
\usepackage{float}
\pdfoutput=1
 \newcommand{\be}{\begin{equation}}
 \newcommand{\ee}{\end{equation}}
 \newcommand{\bea}{\begin{eqnarray}}
 \newcommand{\eea}{\end{eqnarray}}

\newcommand{\beq}{\begin{equation}}
\newcommand{\eeq}{\end{equation}}

\renewcommand*{\thefootnote}{\fnsymbol{footnote}}

\begin{document}

\preprint{IFT-UAM/CSIC-22-156}

\title{Higher-dimensional origin of extended black hole thermodynamics}
\author{Antonia M. Frassino,$^{1}$ Juan F. Pedraza,$^{2}$ Andrew Svesko$^{3}$ and Manus R. Visser$^{4}$}
\affiliation{$^1$Departament de Física Quàntica i Astrofísica and
  Institut de Ciències del Cosmos, Universitat de Barcelona, 08028 Barcelona, Spain\\
 $^2$Instituto de F\'isica Te\'orica UAM/CSIC
Calle Nicol\'as Cabrera 13-15, Cantoblanco, 28049 Madrid, Spain\\
$^3$Department of Physics and Astronomy, University College London, London WC1E 6BT, UK\\
$^4$Department of Applied Mathematics and Theoretical Physics, University of Cambridge, Cambridge CB3 0WA, UK}

\begin{abstract}\vspace{-2mm}
\noindent Holographic braneworlds are used to present a higher-dimensional origin of extended black hole thermodynamics. In this framework, classical, asymptotically anti-de Sitter black holes map to quantum black holes in one dimension less,
with a conformal matter sector that backreacts on the brane geometry. Varying the brane tension alone leads to a dynamical cosmological constant on the brane, and, correspondingly, a variable pressure attributed to the brane black hole. Thus, standard  thermodynamics in the bulk, 
including a work term coming from the brane, induces extended thermodynamics on the brane,
exactly,
to all orders in the backreaction. A microsopic interpretation of the extended thermodynamics of specific quantum black holes is given via double holography.

\end{abstract}

\renewcommand*{\thefootnote}{\arabic{footnote}}
\setcounter{footnote}{0}

\maketitle

\noindent \textbf{Introduction.} Black holes remain an important laboratory for testing ideas in quantum gravity. This is owed to the discovery they behave as thermal systems with an entropy proportional to their horizon area \cite{Bekenstein:1973ur,Hawking:1974sw} 
Though, black holes are peculiar as their thermodynamic first law  lacks a pressure-volume work term. This is because there is no clear notion of pressure or volume for a general black hole. For anti-de Sitter (AdS) black holes  the viewpoint dramatically changes. There, the cosmological constant $\Lambda$ can be identified as pressure, and its variation appears in a generalized first law with a conjugate quantity dubbed the thermodynamic volume 
\cite{Kastor:2009wy,Dolan:2010ha,Cvetic:2010jb}. Thus, AdS black holes give rise to \emph{extended} black hole thermodynamics, offering a rich gravitational perspective on everyday phenomena such as Van der Waals fluids \cite{Chamblin:1999tk,Kubiznak:2012wp}, polymers \cite{Kubiznak:2014zwa}, and heat engines \cite{Johnson:2014yja}.

Despite numerous explorations into extended black hole thermodynamics, there is criticism for identifying the cosmological constant as a varying thermodynamic pressure. A historical inspiration follows from treating $\Lambda$ as a dynamical parameter, leading to additional black hole `hair' \cite{Teitelboim:1985dp}. Further, for consistency, the Smarr relation for AdS black holes requires a pressure-volume term, with pressure proportional to $\Lambda$ \cite{Caldarelli:1999xj,Kastor:2009wy}, however this does not imply that the pressure should be varied. Outside of these considerations, introducing variations of $\Lambda$ is not \emph{a priori} clear from a purely gravitational viewpoint. 

In this letter we provide a new motivation for a varying cosmological constant using holographic braneworlds, where a dynamical $\Lambda$ on a brane naturally arises from tuning the brane tension. In braneworld holography \cite{deHaro:2000wj} one couples a $d$-dimensional brane to a gravitational theory in an asymptotically $(d+1)$-dimensional AdS background, which has a dual interpretation as a  conformal field theory (CFT) living on the asymptotic AdS boundary. Upon integrating out the ultraviolet (UV) degrees of freedom of the CFT, a specific theory of gravity is induced on the brane with an effective cosmological constant controlled by the brane tension. From the higher-dimensional perspective, this tension is a physical tunable parameter, controlling both the coupling to the bulk gravity theory and the position of the brane, leading to variations of the cosmological constant on the brane.

Specific braneworld models are also useful to investigate the microscopic interpretation of extended thermodynamics. To this end, one may consider AdS braneworlds with a second holographic description in terms of a $(d-1)-$dimensional CFT. Via AdS/CFT, variations in $\Lambda$ correspond to variations in the number of degrees of freedom of the dual theory, or the CFT central charge $c$ \cite{Kastor:2009wy,Johnson:2014yja,Dolan:2014cja,Kastor:2014dra,Caceres:2015vsa,Karch:2015rpa}. While AdS/CFT duality is often discussed in the classical  $c\to\infty$ limit, 
it is expected to hold even when all $1/c$ quantum corrections are included.
In fact, going beyond infinite $c$ could be worthwhile since in the classical limit all extensive quantities scale with the number of degrees of freedom \cite{Karch:2015rpa,Visser:2021eqk} and, when properly organized, extended black hole thermodynamics appears to be a ``trivial'' modification of standard thermodynamics. It is only when $1/c$ corrections are included that true new physics could arise, and we can access them via braneworld holography. In this setting, solving classical equations of a braneworld in $(d+1)$-dimensions amounts to solving semi-classical gravitational equations in $d$-dimensions, giving rise to so-called `quantum' black holes on the brane \cite{Emparan:2002px}. Therefore, braneworld holography provides a natural framework to generalize extended thermodynamics of  holographic CFTs to all orders in $1/c$.

\noindent \textbf{Inducing a variation of the  cosmological constant.} Consider an asymptotically $\text{AdS}_{d+1}$ spacetime $\mathcal{M}$ of curvature scale $L_{d+1}$, with a dual description in terms of a $\text{CFT}_{d}$ on the asymptotic boundary $\partial\mathcal{M}$. The quantum fluctuations of the CFT lead to UV divergences which may be removed by adding appropriate counterterms \cite{Kraus:1999di,Papadimitriou:2004ap}, leading to a holographically renormalized theory. 
In braneworld holography \cite{deHaro:2000wj} (see also \cite{Chen:2020uac,Bueno:2022log}), one replaces the regulator surface inside the bulk with a $d$-dimensional brane~$\mathcal{B}$, as in the Randall-Sundrum braneworld construction \cite{Randall:1999vf}, which acts as a physical cutoff and renders the UV divergences finite. Further, the metric on the brane is dynamical, and is governed by a holographically induced higher curvature theory of gravity coupled to a CFT with a UV cutoff.

Precisely, let the bulk be characterized by the action
\be\label{BulkTheory}
I_{\text{Bulk}}[\mathcal{M}]+I_{\text{GHY}}[\partial\mathcal{M}]+I_{\text{Brane}}[\mathcal{B}]\,,
\ee
where
\bea
I_{\text{Bulk}}&=&\frac{1}{16\pi G_{d+1}}\int_{\mathcal{M}} d^{d+1}x\sqrt{-g}\left(\hat{R}-2\Lambda_{d+1}\right)\,,\\
I_{\text{GHY}}&=&\frac{1}{8\pi G_{d+1}}\int_{\partial\mathcal{M}}d^{d}x\sqrt{-h}K\,,\\
I_{\text{Brane}}&=&-\tau \int_{\mathcal{B}}d^{d}x\sqrt{-h}\,.
\eea
Here $\hat{R}$ is the bulk Ricci scalar, 
$\Lambda_{d+1}=-d(d-1)/2L_{d+1}^{2}$ is the cosmological constant, in the Gibbons-Hawking York (GHY) term, $K$ is the trace of the extrinsic curvature of $\partial\mathcal{M}$,
and  $\tau$ is the tension of the brane. 

\begin{figure}[t!]
{\centering
\includegraphics[trim={0.03cm 0 0.02cm 0},clip,width=8.4cm]{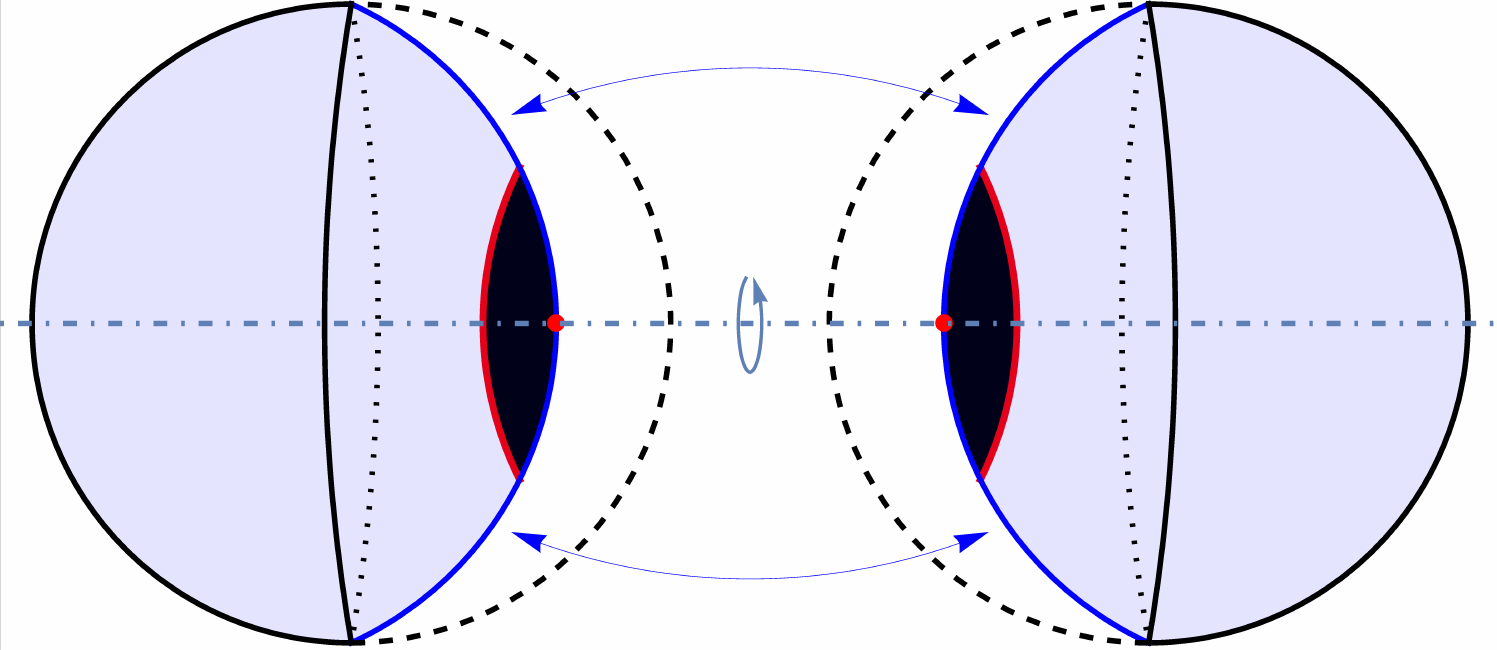}}
\put(-34,80){\footnotesize $\partial\mathcal{M}$}
\put(-48,56){\footnotesize $\mathcal{M}$}
\put(-76,80){\footnotesize {\color{blue}$\mathcal{B}$}}
\put(-169,80){\footnotesize {\color{blue}$\mathcal{B}$}}
\put(-201,56){\footnotesize $\mathcal{M}$}
\put(-221,80){\footnotesize $\partial\mathcal{M}$}
\caption{A $\mathbb{Z}_2$-symmetric, double-sided braneworld construction. The bulk region in white is excised down to the brane $\mathcal{B}$ (blue line), and glued to another copy. A bulk black hole  with event horizon (red line)
is attached to the brane, and induces a horizon on the brane.\vspace{-7mm}
\label{fig:branes}}
\end{figure}
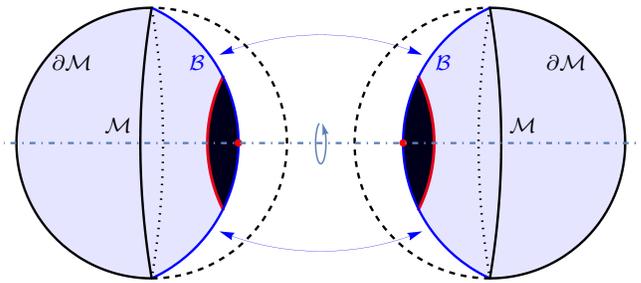

Integrating out the bulk between $\partial\mathcal{M}$ up to $\mathcal{B}$ (typically taken to be near the boundary) amounts to removing CFT degrees of freedom above the UV cutoff, leading to the brane effective action $I$ (see Fig. \ref{fig:branes} for a depiction)
\be
I=I_{\text{Bgrav}}[\mathcal{B}]+I_{\text{CFT}}[\mathcal{B}]\,,
\ee
with the induced gravity theory on the brane
\beq
\begin{split}
I_{\text{Bgrav}}&=\frac{1}{16\pi G_{d}}\int_{\mathcal{B}} d^{d}x\sqrt{-h}\biggr[R-2\Lambda_{d}\\
&+\frac{L_{d+1}^{2}}{(d-4)(d-2)}(R^2\text{-terms})+\cdots\biggr]\,,
\end{split}
\eeq
and two effective parameters endowed from the higher-dimensional parent theory
\bea
G_{d}&=&\frac{d-2}{2L_{d+1}}G_{d+1} \,,\label{eq:effGd}\\
\frac{1}{L_{d}^2}&=&\frac{2}{L_{d+1}^2}\left(1-\frac{4\pi G_{d+1}L_{d+1}}{d-1}\tau\right)\,. \label{eq:effLd}
\eea
Here $L_d$ is the  AdS radius on the brane and
the $R^{2}+...$ terms refer to higher curvature contributions that may be solved for perturbatively, in principle to any order.  The action $I_{\text{CFT}}$ arises from integrating out normalizable modes, 
which accounts for the dual CFT state.

We can understand (\ref{BulkTheory}) as a theory of a finite $(d+1)$-dimensional system with dynamics ruled by general relativity and a brane. Alternatively, from the brane perspective, one has a specific higher-curvature theory in $d$ spacetime dimensions coupled to a cutoff CFT which backreacts on the brane metric. This means classical solutions to the bulk Einstein equations correspond to solutions to the semi-classical equations of motion on the brane. Specifically, classical black holes map to quantum black holes, accounting for all orders of backreaction \cite{Emparan:2002px}.

From the bulk perspective, it is very natural to tune the tension of the brane $\tau$ since it is a physical parameter of the system. The tension determines the position of the brane via Israel junction conditions, thence varying the tension changes the position. 
Accordingly, by virtue of Eq.~(\ref{eq:effLd}), 
varying the tension alone (keeping the other bulk parameters $L_{d+1}$ and $G_{d+1}$ fixed) corresponds to varying the cosmological constant on the brane $\Lambda_{d}$:
\beq
\delta \tau 
=  \frac{\delta \Lambda_d}{8 \pi G_{d}}\,.
\eeq
Notably, this argument holds for general geometric set-ups -- we have not selected particular solutions in the bulk or brane. However, in anticipation of what is to come, it is this observation which leads to our central claim: \emph{classical black hole thermodynamics in the bulk, including work done by the brane, induces extended thermodynamics of quantum black holes on the brane.}

\noindent \textbf{Quantum black holes on the brane.} For an explicit realization of the map between classical and quantum black holes, consider the setup of \cite{Emparan:1999wa,Emparan:1999fd,Emparan:2020znc}, with an $\text{AdS}_{3}$ Karch-Randall brane \cite{Karch:2000ct} in a bulk AdS$_{4}$ C-metric 
\beq 
\begin{split}
ds^{2}&=\frac{\ell^{2}}{(\ell+xr)^{2}}\biggr[-H(r)dt^{2}+\frac{dr^{2}}{H(r)}\\
&+r^{2}\left(\frac{dx^{2}}{G(x)}+G(x)d\phi^{2}\right)\biggr]\,,
\end{split}
\label{eq:Cmet}\eeq
with metric functions
\beq H(r)=\kappa+\frac{r^{2}}{\ell_{3}^{2}}-\frac{\mu\ell}{r}\;,\quad G(x)=1-\kappa x^{2}-\mu x^{3}\;.\label{eq:Hfunc}\eeq
Here $\kappa=\pm1,0$ corresponds to the types of slicings on the brane, where $\kappa=-1$ recovers a classical Banados-Teitelboim-Zanelli (BTZ) black hole \cite{Banados:1992wn,Banados:1992gq} on the brane, but we will leave $\kappa$ arbitrary corresponding to a family of brane black holes.
 The positive parameter $\ell$ is the inverse acceleration associated with the accelerating black hole interpretation of the C-metric and is related to the $\text{AdS}_{4}$ length scale via $L_{4}^{-1}=(\frac{1}{\ell^{2}}+\frac{1}{\ell_{3}^{2}})^{1/2}$, and $\tau$ by
\beq \tau =\frac{1}{2\pi G_4\ell}\,.\label{eq:branetenqBTZ}\eeq
Further,  
$\ell_{3}$ is a length scale related to the curvature scale $L_{3}$ of the $\text{AdS}_{3}$  brane via 
\beq \frac{1}{L_3^2}=\frac{2}{L_{4}^2}\left(1-\frac{L_{4}}{\ell}\right)=\frac{1}{\ell_3^2}\left(1+\frac{\ell^2}{4\ell_3^2}+\mathcal{O}\left(\frac{\ell^{4}}{\ell_{3}^{4}}\right)\right)\,.\eeq
Note it is $L_3$ which appears in the brane action but the solutions are characterized by the AdS radius $\ell_3$, as standard in higher curvature gravities.   Lastly, $\mu$ is a positive parameter related to the mass of the black hole.  

Holographically, excising the bulk region between the $\text{AdS}_{4}$ boundary and the brane (Fig. \ref{fig:branes}) is dual to integrating out UV degrees of freedom of the CFT to a cutoff energy $1/\ell$.
This procedure induces  a quantum BTZ (qBTZ) black hole on the brane at $x=0$ \cite{Emparan:2020znc},
\beq
\begin{split}
&ds^2=-f(r)dt^2+\frac{dr^2}{f(r)}+r^2d\phi^2\,,\\
&f(r)=\frac{r^2}{\ell_3^2}-8\mathcal{G}_3M-\frac{\ell F(M)}{r}\,,
\end{split}
\eeq
where the largest root of $f(r)$ yields the qBTZ horizon $r=r_{+}$. Here $M$ is the black hole mass, $\mathcal{G}_3=G_{3}/\sqrt{1+(\ell/\ell_{3})^{2}}$ is a `renormalized' Newton's constant, and $F(M)$ a form function which can be explicitly determined by solving the semi-classical gravitational equations on the brane. In fact, the renormalized stress-tensor 
of the coupled holographic CFT has the same radial dependence and tensorial structure as the stress tensor of a free massless scalar field conformally coupled to three-dimensional Einstein gravity in the classical BTZ or conical $\text{AdS}_{3}$ backgrounds \cite{Souradeep:1992ia,Steif:1993zv,Casals:2016ioo}, differing only in $F(M)$. 
Notably, the qBTZ black hole is guaranteed to be a solution to the semi-classical theory to all orders in backreaction. 

Thus, $\ell$ controls the strength of  backreaction from the brane perspective. As such, one typically treats $\ell\ll\ell_{3}$, such that induced quantum corrections are perturbatively small compared to the leading order Einstein-Hilbert contribution in the brane effective action. In the bulk this is equivalent to taking the limit where the brane is close to the $\text{AdS}_{4}$ boundary. We emphasize, however, our construction works equally well when $\ell$ is non-zero and finite, where, however, the gravity theory on the brane has a massive graviton \cite{Karch:2000ct}.

The parameter $\ell$ also features into the coupling of the gravity theory to the cutoff CFT on the brane. To see this, note the dictionary for the central charge $c_{3}$ of the holographic $\text{CFT}_{3}$ is \cite{Emparan:2020znc}
\beq c_{3}=\frac{ L^{2}_{4}}{G_{4}}=\frac{\ell}{2G_{3}\sqrt{1+\nu^{2}}}\,,\eeq
where we introduced the positive dimensionless parameter $\nu\equiv\ell/\ell_3$. Even without knowing the explicit form of $I_{\text{CFT}}$, its typical value is $c_{3}$, while the typical value of the gravity action is  $L_{3}/G_{3}$, leading to the dimensionless effective coupling $g_{\text{eff}}=G_{3}c_{3}/L_{3}$ \cite{Emparan:2021hyr}. The central charge is related to the tension via $ \tau\! =\!1/(8\pi  \ell G_3^2 c_{3} )$.
Generically, variations in $\tau$ induce changes in $G_{3}, c_{3}$ and $\ell$, though below we   restrict ourselves to a fixed~$c_{3}$ and~$G_{3}$ ensemble, since that corresponds to keeping $L_4$ and $G_4$ fixed.

\noindent \textbf{Inducing extended thermodynamics.} Despite the fact that the C-metric may be interpreted as an accelerating black hole, it nonetheless has a sensible thermodynamic description. %The solution (10) is static because the black hole is held fixed at a proper distance away from an acceleration horizon. 
Indeed, the C-metric has a time-translation Killing symmetry, which can be used to derive a thermodynamic first law for the black hole horizon. There is no acceleration horizon in our setup,  since we are working in the `small acceleration' regime $\ell>L_{4}$ \cite{Podolsky:2002nk}.

With the brane,
the bulk black hole thermodynamics is identified with the thermodynamics of the quantum black hole. To see this, first consider the case when the tension is fixed.  It is convenient to introduce the dimensionless parameter $z=\ell_3/(r_{+}x_1)\in[0,\infty)$, where $x_{1}$ is the only positive root of $G(x)$ remaining after excising the bulk region at $x=0$. Then, the mass $M$, temperature $T$ and entropy $S$ of the classical bulk black hole are \cite{Emparan:1999fd,Emparan:2020znc}
\bea
M&=&\frac{\sqrt{1+\nu^{2}}}{2G_3}\frac{z^2(1-\nu z^3)(1+\nu z)}{(1+3z^2+2\nu z^3)^2}\,,\\
T&=&\frac{1}{2\pi\ell_3}\frac{z(2+3\nu z+\nu z^3)}{1+3z^2+2\nu z^3}\,,\\
S&=&\frac{\pi \ell_3\sqrt{1+\nu^{2}}}{G_3}\frac{z}{1+3z^2+2\nu z^3}\,.
\eea
Each of these expressions follow from the identification of the free energy with the  Euclidean action of the bulk solution \cite{Kudoh:2004ub}, and, for fixed $\nu$, they obey the first law, $dM=T dS$. From the brane perspective, the classical entropy is identified with the generalized entropy  $S\equiv S_{\text{gen}}$ \cite{Emparan:2006ni,Emparan:2020znc}, and has the same temperature $T$. Thus, the qBTZ first law, valid to all orders in backreaction, is
\beq dM=TdS_{\text{gen}}\,.\label{eq:qbtzfirstlaw}\eeq
A similar relation holds for black holes in semi-classical two-dimensional dilaton gravity \cite{Pedraza:2021cvx,Svesko:2022txo}. 

Let us now consider the case when the brane tension is treated as a thermodynamic variable, similar to  the surface tension of liquids. 
The effect is that the brane will do work on the black hole system, such that the bulk first law becomes 
\be
dM = TdS + A_{\tau} d\tau \,,
\label{eq:firstlawbulk}\ee
where $A_{\tau}\equiv\left(\frac{\partial M}{\partial \tau}\right)_{\hspace{-1mm}S}$ is the variable conjugate to $\tau$ with   scaling dimension  of bulk area. 
We note that the black hole mass plays the role of    enthalpy $H=E+ \tau A_\tau$, since   enthalpy in standard thermodynamics satisfies a first law of the form \eqref{eq:firstlawbulk}, whereas the   internal energy $E$ satisfies the usual first law   $dE = T dS - \tau dA_\tau$. 
Further, the thermodynamic quantities obey a Smarr relation, which we derive in the supplemental material, 
\be
M=2TS-2P_4V_4-  \tau A_\tau\,.
\label{eq:bulksmarr}\ee
This includes a $  \tau A_\tau$ term coming from the brane and the standard $P_4 V_4$ due to the bulk cosmological constant, with `pressure' $P_4=-\Lambda_4/(8\pi G_{4})$,  
and its conjugate `thermodynamic volume' $V_4$. The first law (\ref{eq:firstlawbulk}) corresponds to working in a thermal ensemble of fixed pressure $P_{4}$, such that we fix the bulk scales $L_{4}$ and $G_{4}$. Our approach contrasts previous work on (extended) thermodynamics of accelerating black holes \cite{Appels:2016uha,Anabalon:2018ydc,Anabalon:2018qfv,Gregory:2019dtq, Ball:2020vzo} as we include a brane and work in an ensemble where $P_{4}$ is fixed. These authors also find a mechanical work term similar to ours, though the tension corresponds to that of a cosmic string generating the acceleration of the black hole.

As anticipated, $d\tau$ induces extended thermodynamics on the brane. Particularly, the bulk first law (\ref{eq:firstlawbulk}) maps to  the brane first law
\beq \label{eq:branefirstlaw1} dM=TdS_{\text{gen}}+V_{3}dP_{3} \,,\eeq
generalizing the quantum first law (\ref{eq:qbtzfirstlaw}), where $P_3=-\Lambda_3/(8\pi G_{3})$
is the pressure of the qBTZ black hole and $V_3$ its thermodynamic volume. Moreover, the corresponding Smarr relation of the qBTZ solution also includes a contribution from the central charge $c_{3}$  
\be \label{eq:3dsmarr}
0=TS_{\text{gen}}-2P_3V_3+\mu_3 c_3\,,
\ee
where $\mu_3$ is the chemical potential conjugate variable to the central charge with scaling dimension $[L]^{-1}$. Notably, the mass term is absent from the three-dimensional semi-classical Smarr relation (\ref{eq:3dsmarr}), since $G_3 M$ has vanishing scaling dimension, consistent with the extended thermodynamics of the classical BTZ black hole ($\nu\to0$) \cite{Frassino:2015oca, Frassino:2019fgr}. Further, note that  fixing $L_{4}$ (and $G_4$)  in the bulk has restricted us to the ensemble where $c_{3}$ is kept fixed. The crucial point here is that the variation of the pressure on the brane follows directly from variation of the tension.

We derive exact expressions for the conjugate variables $V_{3}$ and $\mu_3$ in the supplementary material, however, for small backreaction  $\nu\ll1$, we find
\bea
\hspace{-5mm}V_{3}&=&V_{\text{BTZ}} +8\pi \ell_{3}^{2}\frac{z^{3}(1+z^{2})}{(1+3z^{2})^{3}}\,\nu+\ldots,\\
\hspace{-5mm}\mu_3&=&\frac{z^{3}(1-z^{2})}{\ell_{3}(1+3z^{2})^{2}}-\frac{z^{2}(3+9z^{2}+8z^{4}-8z^{6})}{2\ell_{3}(1+3z^{2})^{3}}\nu+\ldots,
\eea
where $V_{\text{BTZ}}$ is the thermodynamic volume of the classical BTZ black hole, when $\nu=0$. 
We see that backreaction of the holographic CFT modifies the classical relation between the volume and entropy, $V_{\text{BTZ}}\sim S_{\text{BTZ}}^{2}$, where $S_{\text{BTZ}}$ is the classical BTZ entropy. Notably, to leading order, the reverse isoperimetric inequality is obeyed~\cite{Cvetic:2010jb}, 
\beq \mathcal{R}\equiv \left(\frac{V_{3}}{\pi}\right)^{1/2}\left(\frac{\pi}{2S_{\text{gen}}}\right)= 1+z\nu+\mathcal{O}(\nu^{2})\geq 1\,.\eeq
Hence, in the limit of small backreaction, the qBTZ is said to be `sub-entropic', i.e., a black hole with volume $V_{3}$ whose entropy is less than a classical BTZ black hole (where $\mathcal{R}=1$) of the same volume.  This inequality is violated, however, for general $z$ and $\nu$. In fact, the volume is bounded above and below, with the lower bound taking a negative value due to a large Casimir effect. A similar feature appears in the mass of the qBTZ solution, which has a finite range and eventually becomes negative as Casimir effects dominate \cite{Emparan:1999fd,Emparan:2020znc}.
These new features are most easily observed in a pressure-volume plot at constant temperature \cite{inprep2}. 

\noindent \textbf{Doubly holographic interpretation.} Since the brane geometry is asymptotically $\text{AdS}_{3}$, a third description of the bulk-brane system emerges. Namely, we may replace the induced gravity theory on the brane with a defect $\text{CFT}_{2}$ coupled to the $\text{CFT}_{3}$, hence, our construction exhibits `double holography' \cite{Almheiri:2019hni}. Per the $\text{AdS}_{3}/\text{CFT}_{2}$ dictionary \cite{Brown:1986nw}, the defect CFT has central charge
\be
c_2=\frac{3L_3}{2G_3}\,.
\ee
For fixed $L_4$ and $G_4$,
we see variations in the brane tension cascade down to variations in $c_{2}$ due to the proportionality with $L_{3}$. This is consistent with the standard microscopic interpretation of a dynamical cosmological constant  \cite{Kastor:2009wy,Johnson:2014yja,Dolan:2014cja,Kastor:2014dra,Caceres:2015vsa,Karch:2015rpa}. 

We can now ascribe a   dual field theory   interpretation to qBTZ thermodynamics. Following \cite{Visser:2021eqk,Cong:2021jgb}, we introduce a conjugate variable to the central charge $c_{2}$, namely, a chemical potential~$\mu_2$. 
We may then   recast the brane extended first law (\ref{eq:firstlawbulk}) in terms of $\text{CFT}_{2}$ quantities
\be \label{CFTfirstlaw}
dE = T dS_2  -P_2dV_{2} + \mu_2 dc_2\,. 
\ee
Here we identified the energy $E$ and entropy $S_{2}$ of the two-dimensional field theory  with $M$ and $S_{\text{gen}}$, respectively.
Further, $V_2 = 2\pi L_3$ and $P_2= M/V_2$ are the volume and pressure in the CFT$_2$. Unlike the brane first law, notice we have a $PdV$ and not $VdP$ term. This suggests the mass $M$ should be understood in the dual CFT as internal energy \cite{Visser:2021eqk}, not enthalpy, as is common in extended black hole thermodynamics \cite{Kastor:2009wy}. By comparing \eqref{eq:branefirstlaw1} and \eqref{CFTfirstlaw} and using the Smarr relation \eqref{eq:3dsmarr} one finds that the chemical potential is proportional to the grand canonical free energy~$W$  
\be
\mu_2 c_2= M -  T S_{\text{gen}} - \mu_3 c_3  \equiv W\,.
\ee
This can be interpreted as the Euler relation in the $\text{CFT}_{2}$, 
\beq E = T S_{2} + \mu_2 c_2 + \mu_n n\,,\eeq
where the coupling $g_{3}$ is dual to a chemical potential $\mu_n$ associated to $c_{3}=n$, the number of the light degrees of freedom that are excited in the $\text{CFT}_{2}$ (corresponding to the cutoff CFT on the brane). Thus, we have found a consistent  thermodynamic field theory description of quantum black holes, and generalized the Euler equation in holographic CFTs to higher order in $1/c.$

\noindent \textbf{Discussion.} We have shown braneworld holography offers a natural framework to explore extended black hole thermodynamics: varying brane tension induces a variable pressure of a quantum black hole on the brane to all orders in $1/c$. As proof of concept, we focused on a specific type of black hole, the qBTZ, exhibiting markedly different behavior from the classical BTZ. In particular, the qBTZ system is sub-entropic, and 
the thermodynamic volume attains a maximum value. 

The investigation here opens up a plethora of future directions. Firstly, our analysis suggests the mass of the bulk is interpreted as an enthalpy,  consistent with the interpretation of black hole mass on the brane. It would be interesting to verify this  using   the on-shell Euclidean action of the bulk black hole, along the lines of \cite{Hawking:1995zn,Kudoh:2004ub,Ball:2020vzo}.  Further, our case study can be extended to other types of quantum black holes, \emph{e.g.}, rotating and charged qBTZ \cite{Emparan:2020znc,inprep}, and quantum de Sitter black holes~\cite{Emparan:2022ijy,Panella:2023lsi}. Extra work terms in these models are expected to enrich the thermodynamic phase structure, for which one may build doubly holographic and quantum heat engines, semi-classical versions of \cite{Johnson:2014yja,Johnson:2019olt}, where universal features of heat engine efficiency \cite{Johnson:2016pfa,DiMarco:2022yhp} will be modified due to backreaction.

Another interesting extension would be to consider more general braneworld models. In our analysis, we assumed a simple brane action with only a tension term. Upon integrating out UV degrees of freedom we thus arrive to a theory where the gravitational strength on the brane and the graviton mass are controlled by the same parameter $\tau$. This in turn imposes an interdependence on various scales in the worldvolume theory, which control the strength of the different (higher curvature) gravity couplings. More generally one may consider situations with induced scale separation. 
One could modify the brane action, for instance, by additional matter fields or via a Dvali-Gabadadze-Porrati (DGP) term \cite{Dvali:2000hr}, \emph{e.g.},
%For example, consider the modified brane action
\beq I_{\text{Brane}}=\int_{\mathcal{B}}d^{d}x\sqrt{-h}\left(-\tau%-\Delta\tau)
+\frac{1}{16\pi G_{\mathcal{B}}}R\right)\,.\eeq
The DGP term separates the graviton mass from the gravitational strength on the brane, and has proven insightful for calculations of semi-classical entropy \cite{Chen:2020uac,Chen:2020hmv}. Generally, the separation of scales will be induced because the Einstein-Hilbert term will change the position of the brane by altering the brane stress-tensor and also modify the effective brane cosmological constant. Alternately, by tuning the tension such that the brane position is unchanged, the Newton's constant on the brane $G_{\mathcal{B}}$ directly influences the effective Newton's constant via \cite{Chen:2020uac}
\beq G_{d}^{-1}=2L_{d+1}/((d-2)G_{d+1})+G_{\mathcal{B}}^{-1}\,,\eeq
such that $\delta G_{\mathcal{B}}/G_{\mathcal{B}}^{2}=\delta G_{d}/G_{d}^{2}$.   
Thus, even if the gravitational parameters are held fixed in the bulk, the DGP term could allow for richer induced brane thermodynamics. The DGP term induces a variation of Newton's constant, which in turn yields a variation of the central charge $c_3$,  kept fixed in this paper \cite{Karch:2015rpa,Visser:2021eqk,Cong:2021fnf}.
One may also consider including topological gravitational terms, such as Gauss–Bonnet or Lovelock densities, or theories coupled with matter, all of which introduce additional tunable parameters. Adding such DGP-like terms  allows one to study extended first laws of entanglement in the presence of additional central charges, as in \cite{Kastor:2014dra,Caceres:2016xjz,Johnson:2018amj,Rosso:2020zkk}, but beyond their classical limits.

Lastly, while we have provided a higher-dimensional origin of extended thermodynamics using a bottom-up construction,
it is desired to have a more fundamental string theoretic underpinning. Doing so would grant us access to stringy corrections beyond the semi-classical regime considered here. Such a program has been initiated in \cite{Meessen:2022hcg}. %Connections to string theory  have been made in \cite{Meessen:2022hcg}, where the cosmological constant is promoted to a form potential and fits in the embedding tensor in gauged supergravity, the components of which typically arise from fluxes in the higher dimensional theory. 
It is also worth constructing braneworld black holes in string or M-theory and studying their doubly holographic description, in the spirit of \cite{Karch:2000gx,Karch:2001cw,DHoker:2007zhm,Aharony:2011yc,Karch:2022rvr}. 
%string theory constructions of braneworlds and their doubly holographic descriptions \cite{Karch:2022rvr}. 

\noindent \emph{Acknowledgements.}
We are grateful to Roberto Emparan, Robie Hennigar and Clifford Johnson for useful discussions and correspondence.  AMF is supported by ERC Advanced Grant GravBHs-692951, MICINN
grant PID2019-105614GB-C22, AGAUR grant 2017-SGR 754, and State Research Agency of
MICINN through the “Unit of Excellence Mar\'ia de Maeztu 2020-2023” award to the Institute of Cosmos Sciences (CEX2019-000918-M). JFP is supported by the `Atracci\'on de Talento' program grant 2020-T1/TIC-20495 and by the Spanish Research Agency through the grants CEX2020-001007-S and PID2021-123017NB-I00, funded by MCIN/AEI/10.13039/501100011033 and by ERDF A way of making Europe. AS is supported by the Simons Foundation via \emph{It from Qubit Collaboration} and by EPSRC. MRV is supported by  SNF Postdoc Mobility grant P500PT-206877 ``Semi-classical thermodynamics of black holes and the information paradox''.

\bibliographystyle{apsrev4-2}
\bibliography{qdSrefs}

\newpage

\begin{widetext}
\section{Supplementary material}

\subsection{Generalized  Smarr relations between thermodynamic quantities}

Below we derive the generalized Smarr relations and list the exact, extended thermodynamic quantities for the qBTZ and the bulk black hole with a codimension-1 brane of tension $\tau$.
%The importance of the Smarr relations goes beyond mere algebraic relations  \cite{Meessen:2022hcg}.
Let us begin with the bulk black hole plus brane configuration. By a scaling argument analogous to \cite{Kastor:2009wy}, we can write down a bulk Smarr relation (including a tension contribution, similar to \cite{Armas:2015qsv}). 
To this end, recall that in $D$-spacetime dimensions, $[G_{D}]=[L]^{D-2}=[A_{D}]$, $[M]=[L]^{-1}$, $[\Lambda_{D}]=[L]^{-2}$, and $[\tau]=[L]^{-(D-1)}$, such that $[G_{D}M]=[L]^{D-3}$,  and $[G_{D}\tau]=[L]^{-1}$. Restricting to $D=4$, we express the product $ G_4 M$ as the homogeneous function  $G_4 M = \mathcal M  (A_4, \Lambda_4, G_4 \tau)$, where $A_{4}$ is the horizon area of the four-dimensional black hole.
Euler's theorem for  homogeneous functions then implies
\beq G_{4}M=2\left(\frac{\partial (G_4 M)}{\partial A_{4}}\right)_{\hspace{-1mm} \Lambda_{4},\tau}A_{4}-2\left(\frac{\partial (G_4 M)}{\partial \Lambda_{4}}\right)_{\hspace{-1mm} A_{4},\tau}\Lambda_{4}-\left(\frac{\partial M}{\partial \tau}\right)_{\hspace{-1mm} A_{4},\Lambda_{4}}G_4 \tau\,.\eeq
Upon identifying $P_{4}=-\Lambda_{4}/8\pi G_{4}$, and $S=A_{4}/4G_{4}$, we are led to the four-dimensional Smarr law
\beq
M=2TS-2P_4V_4-  \tau A_\tau\,,
\eeq
 where we have defined
conjugate variables $V_{4}$ and $A_{\tau}$ as follows
\beq V_{4}\equiv \left(\frac{\partial M}{\partial P_{4}}\right)_{\hspace{-1mm}S,\tau}=\frac{8\pi \ell_{3}^{3}}{3}\frac{z^{2}\nu(1+2z\nu)}{(1+3z^{2}+2\nu z^{3})^{2}}
\,.\eeq
\beq A_{\tau}\equiv \left(\frac{\partial M}{\partial \tau}\right)_{\hspace{-1mm}S,P_{4}}=\frac{2\pi \ell_{3}^{2} z^{2}[-2+\nu^{2}+3\nu^{3}z^{3}+\nu^{4}z^{4}+\nu z(\nu^{2}-4)]}{(1+3z^{2}+2\nu z^{3})^{2}}\,.\label{eq:Atausupp}\eeq
In the second equality we  computed the conjugate variables explicitly as a function of $\ell_3$, $z=\ell_3/(r_+ x_1)$ and $\nu =\ell / \ell_3$. The quantity  $V_{4}$ may be understood as the thermodynamic volume of the bulk black hole.  In the large $\ell$ limit, where the brane becomes tensionless and the bulk black hole has vanishing acceleration, we recover the geometric volume of an $\text{AdS}_{4}$-Schwarzschild black hole, while $A_{\tau}$ diverges as $\pi \ell^{2}_{3}\nu^{2}/2$.

The Smarr formula on the brane follows a similar scaling argument, though is a little more subtle. First recall black hole thermodynamics in the bulk induces black hole thermodynamics on the brane, such that, from the brane perspective, $S=S_{\text{gen}}$ and $M$ is the mass of the black hole \cite{Emparan:1999fd}.    Further, the relevant scales on the brane are: $L_{3}$ (or $\Lambda_{3}$), $G_3 c_{3}$ and $G_{3}M$, such that in three spacetime dimensions $G_3 M$ has a vanishing scaling dimension, while $G_{3}c_3$ has scaling dimension $+1$ and $\Lambda_{3}$ has dimension $-2$. Moreover, the three-dimensional generalized entropy times Newton's constant has a scaling dimension of $+1$ since  it can be expressed as $ G_3 S_{\text{gen}}=A_{3}f(\nu,z)$, where $A_{3}$ is the area of the three-dimensional black hole horizon, and $f(\nu,z)$ is a dimensionless function. Therefore, we may now cast the mass as the homogeneous function, $G_3 M=\mathcal M(A_{3},\Lambda_{3},G_{3}c_3)$, such that via   Euler's theorem we have
\beq 0=\left(\frac{\partial (G_3M)}{\partial A_{3}}\right)_{\hspace{-1mm}\Lambda_{3},c_{3}}A_{3}-2\left(\frac{\partial (G_3 M)}{\partial \Lambda_{3}}\right)_{\hspace{-1mm} A_{3},c_{3}}\Lambda_{3}+\left(\frac{\partial M}{\partial c_{3}}\right)_{\hspace{-1mm} A_{3},\Lambda_{3}}G_{3}c_3\,.\eeq
When we identify $P_{3}=-\Lambda_{3}/8\pi G_{3}$ and $S_{\text{gen}}=A_{3}f(\nu,z)/G_3$, we uncover the three-dimensional Smarr law
\beq 0=TS_{\text{gen}}-2P_{3}V_{3}+\mu_3 c_{3}\,,\eeq
where the conjugate terms $V_{3}$ and $\mu_{3}$ are defined as
\beq V_{3}\equiv \left(\frac{\partial M}{\partial P_{3}}\right)_{\hspace{-1mm}S_{\text{gen}},c_{3}}=-\frac{2\pi\ell_{3}^{2}z^{2}[-2+\nu^{2}+3\nu^{3}z^{3}+\nu^{4} z^{4} +\nu z(\nu^{2}-4)]}{(1+3z^{2}+2\nu z^{3})^{2}}\,,\eeq 
\beq \mu_3\equiv \left(\frac{\partial M}{\partial c_{3}}\right)_{\hspace{-1mm}S_{\text{gen}},P_{3}}\hspace{-1mm}=-\frac{z^{2}(1+\nu^{2})\{\nu^{2}(2+\nu z(3+z^{2}))+2(\sqrt{1+\nu^{2}}-1)[\nu^{2}-2+z\nu(\nu^{2}-4+\nu^{2}z^{2}(3+z\nu))]\}}{\ell_{3}\nu^{3}(1+3z^{2}+2\nu z^{3})^{2}}\;. \eeq
These are \emph{exact} relations as no assumption about the relative size between $\ell$ and $\ell_{3}$ was made. Notice that the thermodynamic volume on the brane equals \emph{minus} the variable $A_{\tau}$ (\ref{eq:Atausupp}).

\end{widetext}

\end{document}